	\documentclass[12pt,table]{article}
	
	\usepackage[top=1.5in, bottom=1.5in, left= 1.0in, right=1.0in]{geometry}

	\usepackage[margin= 16pt , footnotesize ,  sc , labelfont=bf , labelsep=period , format=plain , justification=centerfirst]{caption}
	\usepackage{floatrow}
		\usepackage{listings}
		\usepackage{amsfonts}
		\usepackage{amsmath}
		\usepackage{amssymb}
		\usepackage{amsthm}
		\usepackage{array}
		\usepackage{booktabs,longtable}
		\usepackage{eurosym}
		\usepackage{floatrow}
		\usepackage{graphicx}
		\usepackage{hyperref}
		\usepackage{mathrsfs}

		\usepackage{color, colortbl}
		\usepackage[english]{babel}
		\usepackage{multirow}
		\usepackage{lscape}
		\usepackage{pdflscape}
		\usepackage{rotating}
		\usepackage{subfig}
		\usepackage{setspace}
		\usepackage{titlesec}
		\usepackage{verbatim}
		\usepackage[pdftex,dvipsnames]{xcolor}
		\usepackage[textsize=tiny]{todonotes}
		\usepackage{array}
		\newcolumntype{C}[1]{>{\centering\let\newline\\\arraybackslash\hspace{0pt}}m{#1}}

		\usepackage{natbib}
		\setcitestyle{authoryear,open={(},close={)}}

	\usepackage{tabularx, booktabs}
	\newcolumntype{Y}{>{\centering\arraybackslash}X}

		\makeatletter
		
		\newcommand{\Rmnum}[1]{\expandafter\@slowromancap\romannumeral #1@}
		\makeatother

	\makeatletter
	\pdfpageheight\paperheight
	\pdfpagewidth\paperwidth
	\usepackage{csquotes}
	\usepackage{subfig}
	\usepackage{ragged2e}
	\usepackage[hang,flushmargin]{footmisc}

	\newcommand{\tstats}{\textit{T}-stats are reported in parentheses. Asterisks denote significance levels (***$ =1\%$, **$ =5\%$, *$ =10\%$).}
	\newcommand{\postwin}{$[1,24]$}
	\newcommand{\prewin}{ $[-24,0]$}
	
	\definecolor{electric_blue}{HTML}{023EFF}
	\definecolor{royal_blue}{HTML}{0071BC}
	\definecolor{navy_blue}{HTML}{006EB8}
	\definecolor{midnight_blue}{HTML}{191E78}
	\definecolor{dark_grey}{HTML}{666666}
	
	\usepackage{xcolor}
    	\usepackage{sectsty}
	\sectionfont{\color{midnight_blue}}  
	\subsectionfont{\color{midnight_blue}}
	\subsubsectionfont{\color{midnight_blue}}
	
	\hypersetup{
        colorlinks,
        linkcolor={electric_blue},
        citecolor={electric_blue},
        urlcolor={electric_blue}
    }

\newcommand{\figurecap}[4]{ 
	\begin{figure}[t]
		\centering
		\begin{minipage}[c]{1\textwidth}
			\begin{centering}
				\vspace{-1em}
				\includegraphics[width=#3\linewidth]{IMG/#2}
				\captionsetup{justification=justified, singlelinecheck=true, font=footnotesize} 
				\caption{\textbf{#1.} #4 \vspace{1em}}
				\label{fig:#2}
			\end{centering}
		\end{minipage}
	\end{figure}
}

\usepackage{fancyvrb,xcolor}

\begin{document}

		\title{
			\Large{ \bf NFT Bubbles }
			\\\vspace{0.3em}
		}

\date{March 2023}
			\author{Andrea Barbon and Angelo Ranaldo\thanks{\fontsize{9}{9}\selectfont 
			Andrea Barbon (\href{mailto:andrea.barbon@unisg.ch}{andrea.barbon@unisg.ch}) and Angelo Ranaldo (\href{mailto:angelo.ranaldo@unisg.ch}{angelo.ranaldo@unisg.ch}) are both at the University of St. Gallen and Swiss Finance Institute. Andrea Barbon is the corresponding author. We thank Kirill Kazakov for the excellent research assistance. We acknowledge financial support from the Swiss National Science Foundation (SNSF grant 204721).
			}
		}

		\maketitle
		\vspace{-1cm}
		\begin{abstract}

			By investigating nonfungible tokens (NFTs), we provide the first systematic study of retail investor behavior through asset bubbles. 
			Given that NFTs are recorded in public blockchains, we are able to track investor behavior over time, leading to the identification of numerous price run-ups and crashes. 
			Our study reveals that agent-level variables, such as investor sophistication, heterogeneity, and wash trading, in addition to aggregate variables, such as volatility, price acceleration, and turnover, significantly predict bubble formation and price crashes. 
			We find that sophisticated investors consistently outperform others and exhibit characteristics consistent with superior information and skills, supporting the narrative surrounding asset pricing bubbles.
			%
			%
		\end{abstract}
		
		\thispagestyle{empty} \newpage

		``\textit{Speculators may do no harm as bubbles on a steady stream of enterprise. But the position is serious when enterprise becomes the bubble on a whirlpool of speculation}." (\cite{Keynes36}, p. 101)
	 	
    \section{Introduction}\label{sec:intro}
        Ever since Keynes, the narrative on speculative bubbles has contemplated two types of investors: pure speculators, who are mainly interested in the resale price of financial securities, and more sophisticated investors, who assess the market price compared to the fundamental value. The prevalence of speculators over sophisticated investors is the breeding ground for asset bubbles \citep{Kindleberger1978}. Although intuitive, the literature has not yet shed full light on this idea, and it has not yet been empirically ascertained which types of investors contribute to the formation and bursting of asset bubbles. One possible reason for this is the difficulty of tracing investor behavior through speculative bubbles.

        This paper provides the first systematic study of retail investor behavior through asset bubbles. To do this, we exploit the distinctive features of nonfungible tokens (NFTs), which are unique digital assets providing proof of ownership and verification of authenticity held in the blockchain. There are at least three good reasons for leveraging the NFT market for this research purpose. 
        First, transaction attributes and traders' identities are highly transparent since every transaction is recorded on a public blockchain. 
        Second, even though they were introduced only a few years ago,\footnote{The first NFT (Quantum) was created by Kevin McCoy and Anil Dash in May 2014 \citep{Creighton.2022}.} the volume traded in NFTs as of February 2023 is roughly \$73 billion.\footnote{\cite{Barbon2023} explain the crucial aspects of the NFT and its market as well as quantify its size by considering the largest NFT exchanges, which are OpenSea, SuperRare, LooksRare, Foundation, and Rarible. 
        } 
        In this market populated only by retail investors, a legitimate question that has been asked by many is whether it has been a ``balloon" blown up by a technological innovation \citep{vanhorne1985} or a real speculative bubble. Third, it is more difficult even for a sophisticated investor to determine the fundamental value of these assets and arbitrage their deviations from market prices given their uniqueness (or non-fungibility), the absence of cash flow streams, and the inability to sell them short.
        Short selling NFTs in strictly nonfungible collections poses significant challenges due to the unique nature of each token. Covering a short position is difficult as the short seller must acquire the exact same token, which may result in a short squeeze. While short selling may be feasible for semifungible collections with large supplies, this analysis focuses on strictly non-fungible collections for consistency.
        All these aspects offer a unique opportunity to study how the behavior of retail investors relates to the formation and burst of asset bubbles.

        The key question that motivates our research is whether the predictive power of asset bubbles is more effectively discerned by scrutinizing the conduct of individual investors. In line with the asset bubble narrative, we posit that if there exists a larger cohort of sophisticated investors with a heightened degree of skill and expertise, their augmented presence should predict fewer incidents of speculative bubble bursts. Three main results stand out from our analysis: First, aggregate variables such as high volatility, price acceleration, and low turnover significantly predict crashes. Second, some investors persistently perform better than others, and their characteristics suggest that they are more sophisticated, with superior information and skills. Third, a greater presence of sophisticated agents in the price run-up significantly decreases the risk of a subsequent crash and increases ex-post positive returns. In contrast, a smaller number of individual investors and more price manipulation through ``wash trading" increase the crash risk and predict more illiquidity in the crash phase, thus exacerbating the adverse effects.
        
        %

        We proceed in three steps. First, we build a representative data set of the entire NFT market, including the 1,000 most traded NFT collections\footnote{
        An NFT collection is a set of NFT tokens based on the same smart contract. Usually, they are produced by the same artist and share a common theme. For instance, the \href{https://opensea.io/collection/boredapeyachtclub}{Bored Ape Yacht Club} or the \href{https://opensea.io/collection/cryptopunks}{CryptoPunks} are examples of collections.
        } on the largest NFT marketplace, \href{https://www.opensea.io}{OpenSea}. Over the sample period from January 2021 to September 2022, this data set includes more than 15 million transactions from 1.3 million unique wallets, generating a total trading volume of more than 18 billion USD (8.5 million ETH). To identify run-up and crash phases, we adapt the methodology of \cite{greenwood2019bubbles} to determine high-frequency price movements that can be considered short-term bubbles in the same vein as the dot-com bubble literature \citep[e.g.,][]{Ofek2003,Brunnermeier2004}. We identify around 1,000 price run-up events, half of which turn into ex-post crashes. The first question we address is whether crashes can be predicted using aggregate variables. We find strong empirical evidence that collection-specific variables, rather than market-wide variables, predict price crashes. More precisely, the higher the volatility and acceleration of the price, the more likely it is to crash. On the contrary, high turnover is associated with a lower probability of crashing, suggesting that, due to an inability to sell short, a lower concentration of optimistic investors \citep{Scheinkman2003} and smaller liquidity risk predict smoother ex-post price adjustments \citep{Nneji2015}. 
        The picture that shines through this first analysis is that NFT bubbles are predictable, and they are not very dissimilar to those in equities since they share similar asset-specific predictors of price crashes.

        As a second step, we turn to the next question that guides our research: Are there more sophisticated retail investors? If so, what role do they play in the formation and bursting of asset bubbles? A priori, it is not obvious that there may be sophisticated investors in this market. On the one hand, this market is populated only by retail investors rather than institutional investors or hedge funds that presumably have more advanced skills \citep{Brunnermeier2004}. On the other hand, the fundamental value of these new digital assets is more challenging to determine than in other financial securities given the absence of expected cash flow streams originated, for example, by dividends or coupon payments, and that their uniqueness (or nonfungibility) makes it arduous to find close substitutes for hedging or arbitrage purposes. In addition, an investor who believes the price is excessive relative to the fundamental value cannot sell short in this market. Against this background, we ascertain whether there is heterogeneity among investors, whether that heterogeneity is persistent, and whether it can explain investor performance. 
        %
        
        Two main findings emerge: First, the agent's profitability is persistent across run-up events, suggesting that some agents possess more advanced abilities to perform consistently better over time. 
        Second, outperforming agents are distinguished in terms of (a) greater market presence, (b) financial leverage, and (c) sharper timing. In addition to trading more volume and frequently, their activity spans multiple exchanges (cross-market trading), they are more likely to provide liquidity and to trade in decentralized exchanges (DEX), like Uniswap, Balancer, or Curve, and they leverage up their positions to exploit profit opportunities and access arbitrage capital through decentralized lending platforms. Furthermore, these investors are more skilled in timing the bubbles, by both buying early into the run-up phase and selling close to the peak of the bubble. 
        %
        %
        All in all, the more sophisticated behavior of outperforming investors suggests that some individuals possess superior information and skills in identifying NFT investment opportunities, timing the market movements, and avoiding being caught in the bursting of NFT bubbles. 

        Our study brings forth a third innovation by demonstrating that, in addition to (collection-specific) aggregate variables, agent-level variables can assist in predicting asset price crashes and ex-post returns. 
        Specifically, we include three metrics that are measured ex-ante as additional predictive variables for crashes and find that the prediction accuracy improves.
        First, we provide compelling evidence that a greater presence of sophisticated investors is negatively related to ex-post crashes and positively related to ex-post returns, consistent with the narrative about asset price bubbles and the idea that sophisticated investors have superior skills in choosing profitable investments less prone to overvaluation and doing so with good timing. Rather than a causal interpretation, these results may simply highlight the ability of sophisticated agents to estimate ex-ante the probability of ex-post crashes and, thus, self-select into noncrash events.
        Second, we calculate a measure of ownership concentration based on the number of individual investors owning an NFT of the collection relative to the available supply.
        We find that an increase in individual owners predicts lower crash risk and greater liquidity in the ex-post price correction phase. These results support our hypothesis that wider heterogeneity reduces the risk of concentration in optimistic investors and mitigates the adverse effects of the short sale impossibility. It also reduces liquidity risk in future transactions as a larger and more diverse community of traders facilitates resale opportunities, reduces search costs, and increases risk sharing. 
        
        Third, we analyze wash trading\footnote{We broadly define wash trading following the Commodity Futures Trading Commission (CFTC) concept: ``transactions to give the appearance that purchases and sales have been made, without incurring market risk or changing the trader's market position" (CFTC, 2022).} at the aggregate and transaction levels, identifying suspicious trades empirically. Although this practice of price manipulation is not widespread in OpenSea and in our data set, we find that when wash trading is more pervasive ex-ante, ex-post crashes and liquidity deterioration are more likely.

        Our paper contributes to two strands of the literature. First, we add to an extensive literature on asset bubbles.\footnote{See \cite{Brunnermeier2016} for a recent survey of this literature.} Some papers suggest that bubbles could be driven by asymmetric information \citep[e.g.,][]{ALLEN1993206}, limit to arbitrage \citep[e.g.,][]{https://doi.org/10.1111/j.1540-6261.1997.tb03807.x}, heterogeneous beliefs and individual characteristics \citep[e.g.,][]{abreu2003bubbles,https://doi.org/10.1111/0022-1082.00184,Scheinkman2003}, and risk premia combined with learning about new technologies \citep[e.g.,][]{Pastor2009}. We extend the previous literature by empirically analyzing the run-up and crash of asset bubbles on an \textit{agent} basis. The NFT market provides us with an ideal laboratory for analyzing a large number of episodes subject to the impossibility of short selling and accurately identifying individual retail investors. In contrast to prior research, we do not need to conjecture the investor's sophistication by contrasting institutional and retail investors \citep{Ofek2003} or imposing a priori that sophistication pertains only to hedge fund managers \citep{Brunnermeier2004}. In doing so, we highlight new characteristics of individual investors who outperform in bubble episodes such as advanced skills in timing, leveraging, and cross-market activities. Our results suggest that (i) rather than pure speculators, outperforming investors are more sophisticated in terms of information and skills, and (ii) a larger share of sophisticated investors is negatively (positively) related to bubble bursts (ex-post returns), a result so far only documented with experimental approaches \citep{bosch2018cognitive}.

        Second, we add to the nascent literature on blockchain-based assets. So far, only a few studies on NFT have been carried out. There are three notable exceptions related to our work: First, \cite{borri2022economics}, who construct an NFT market index and conduct an asset pricing analysis. Second, \cite{Ito2022} conduct a time-series analysis of NFT prices that can turn into bubbles using logarithmic periodic power law (LPPL) models. Third, \cite{Oh2022} analyze whether investors’ experience predicts higher NFT returns. Other papers focus on specific NFTs such as (digital) real estate in Decentraland \citep{Goldberg2021,DOWLING2022102096}, CryptoKitties \citep{Kireyev2021}, and CryptoPunks \citep{Kong2021}. \cite{Nadini2022} focus on the investor's propensity to buy similar NFTs. Other papers deal with the empirical features of NFT prices. For example, \cite{DOWLING2022102097} analyzes the relation between NFT and cryptocurrency prices, \cite{YOUSAF2022103175} examine return-volume and return-volatility connectedness in NFTs, and \cite{Ante2022} examines spillover effects among NFT sales and between NFT and cryptocurrency prices using error correction models. 
        We add to the NFT literature by investigating the individual characteristics of NFT investors and how these predict aggregate NFT price movements. 
        Since those who invest in NFTs are typically retail investors rather than traditional financial institutions, our work can also be considered a contribution to the FinTech and household finance literature \citep{https://doi.org/10.1111/j.1540-6261.2006.00883.x}. 

        The paper is organized as follows. Section \ref{sec:data} presents the data. Section \ref{sec:bubbles} identify price run-ups and crashes. Section \ref{sec:agents} analyzes individual investors. Section \ref{sec:conclusion} concludes.

	\section{Data and Variables}\label{sec:data}
        We construct a representative data set for the total NFT market by downloading transaction-level data for the 1,000 most traded NFT collections on the largest NFT marketplace, \href{https://www.opensea.io}{OpenSea}.\footnote{OpenSea is a decentralized marketplace for buying, selling, and creating NFTs on the Ethereum blockchain. It provides a platform for creators and users to trade and showcase their virtual assets, such as digital art, gaming items, and more, in a secure and transparent manner.} The choice of collections is based on the total volume traded (denominated in ETH) over the entire lifespan of each collection.
        We limit the sample to collections based on smart contracts complying with the ERC-721 standard,
        thus excluding from the sample semifungible collections.\footnote{
        	The \href{https://eips.ethereum.org/EIPS/eip-721}{Ethereum Request for Comments (ERC) 721} is a standard for smart contracts operating nonfungible tokens. 
        	There exist more general standards, for instance, the \href{https://eips.ethereum.org/EIPS/eip-1155}{ERC-1155}, operating both non-fungible and semifungible tokens.
        } 
        This provides us with a homogeneous universe of collections in which single NFTs are practically impossible to sell short.
        The sample period ranges from January 2021 to September 2022, covering 621 trading days. 
        The data include roughly 15 million transactions from 1.3 million unique wallets, generating a total trading volume of about 20 billion USD (9 million ETH). 
        The average daily volume is 11.5 million USD.
        According to data from Dune Analytics, our sample covers around 5\% of the total dollar volume of NFTs across all exchanges and all blockchains over the same historical period.
        We use average transaction prices to construct a panel at the hourly frequency consisting of 3,535,662 observations. The panel includes the following list of variables measured at the collection hour level:
        \\[0.5em]\textit{Price:} Average of sale prices, expressed in ETH.
    	\\\textit{Floor:} Lowest price at which an NFT of the collection is listed for sale.
    	\\\textit{Return:} Percentage change in price.
    	\\\textit{Volume:} Trading volume, expressed in ETH.
    	\\\textit{Sales:} Count of the number of NFTs sold.
    	\\\textit{Minted:} Number of NFTs minted from the contract, purchased or airdropped.
    	\\\textit{Supply:} Number of NFTs available in the collection (cumulative sum of mints).
    	\\\textit{Turnover:} Ratio of sales divided by supply, expressed in percentage terms.
    	\\\textit{Market Captitalization:} Product of supply and floor price.
    	\\\textit{Age:} Hours elapsed from the creation of the collection.
    	\\[1em]
        %
        %
		The findings of our study regarding the average transaction price and total trading volume of NFT collections are presented in Figure \ref{fig:NFT_market}. To provide a detailed overview of the sample data, we report the summary statistics of the panel data, which have been winsorized at the 1\% level, in Table \ref{tab:summary_stats}. As illustrated in the table, the average price of an NFT in our sample is approximately 0.92 ETH, equivalent to roughly 2,000 USD. Additionally, the average market capitalization is 10 million USD. These relatively high valuations are consistent with our focus on collections that have generated significant trading volume.
		Additionally, the hourly percentage price changes, on average, are higher than 5\%, but with a wide dispersion, reaching 44\% volatility. 
		The mean turnover ratio is 0.33\% at the hourly frequency, implying that, on average, 8\% of the outstanding NFTs of a collection are traded on any given day. 
		Finally, the average time elapsed since the creation of a collection is 114 days, which indicates that most collections are present during a large portion of our sample period.

	\begin{table}[t!]
		\centering
		\resizebox{0.94\textwidth}{!}{\begin{tabular}{lrrrrrr}
\toprule
                   &$\quad$ Count  $\quad$   &$\quad$ Mean  $\quad$   &$\quad$ Std  $\quad$     &$\quad$ 10\%$\quad$   &$\quad$ 50\%$\quad$     &$\quad$ 90\%$\quad$      \\
\midrule
\textit{Price} ($\Xi$)      & 3,511,676 & 0.92     & 2.19      & 0.04   & 0.22     & 2.01      \\
\textit{Floor} ($\Xi$)      & 3,534,040 & 0.80     & 2.05      & 0.03   & 0.18     & 1.73      \\
\textit{Return} (\%)        & 3,508,406 & 5.63     & 44.04     & -16.46 & 0.00     & 19.62     \\
\textit{Volume} ($\Xi$)     & 3,535,662 & 1.24     & 4.42      & 0.00   & 0.00     & 2.38      \\
\textit{Sales}              & 3,535,662 & 1.66     & 4.42      & 0.00   & 0.00     & 4.00      \\
\textit{Minted}             & 3,535,662 & 8,102.61 & 7,998.05  & 623.00 & 7,795.00 & 11,312.00 \\
\textit{Supply}             & 3,440,668 & 6,467.23 & 9,027.44  & 341.00 & 3,974.00 & 11,111.00 \\
\textit{Turnover} (\%)      & 3,439,122 & 0.33     & 1.34      & 0.00   & 0.02     & 0.46      \\
\textit{Market Cap} ($\Xi$) & 3,534,040 & 5,254.65 & 15,264.94 & 70.83  & 1,100.31 & 9,745.52  \\
\textit{Age (hours)}        & 3,535,662 & 2,739.42 & 2,095.13  & 443.00 & 2,307.00 & 5,648.00  \\
\bottomrule
\end{tabular}}
		\captionsetup{justification=justified, singlelinecheck=on, font=footnotesize}
		\caption{\textbf{Summary Statistics.} 
        	The table provides summary statistics for the variables contained in our panel at the collection-hour level.
        	An NFT collection is a set of NFT tokens based on the same smart contract, usually produced by the same artist, and sharing a common theme. The variables include: 
        	\emph{Price}, the average of sale prices, expressed in ETH;
			\emph{Floor}, the lowest price at which an NFT of the collection is listed for sale;
			\emph{Return}, the percentage change in price;
			\emph{Volume}, the trading volume expressed in ETH;
			\emph{Sales}, the count of the number of NFTs sold;
			\emph{Minted}, the number of NFTs minted from the contract, purchased, or airdropped;
			\emph{Supply}, the number of NFTs available in the collection (cumulative sum of mints);
			\emph{Turnover}, the ratio of sales divided by supply, expressed as a percentage;
			\emph{Market Capitalization}, the product of supply and floor price;
			\emph{Age}, the hours elapsed from the creation of the collection.
			All variables are winsorized at the 1\% level.
        }\label{tab:summary_stats}
		\vspace{0.25em}
	\end{table}

        \figurecap{NFT market}{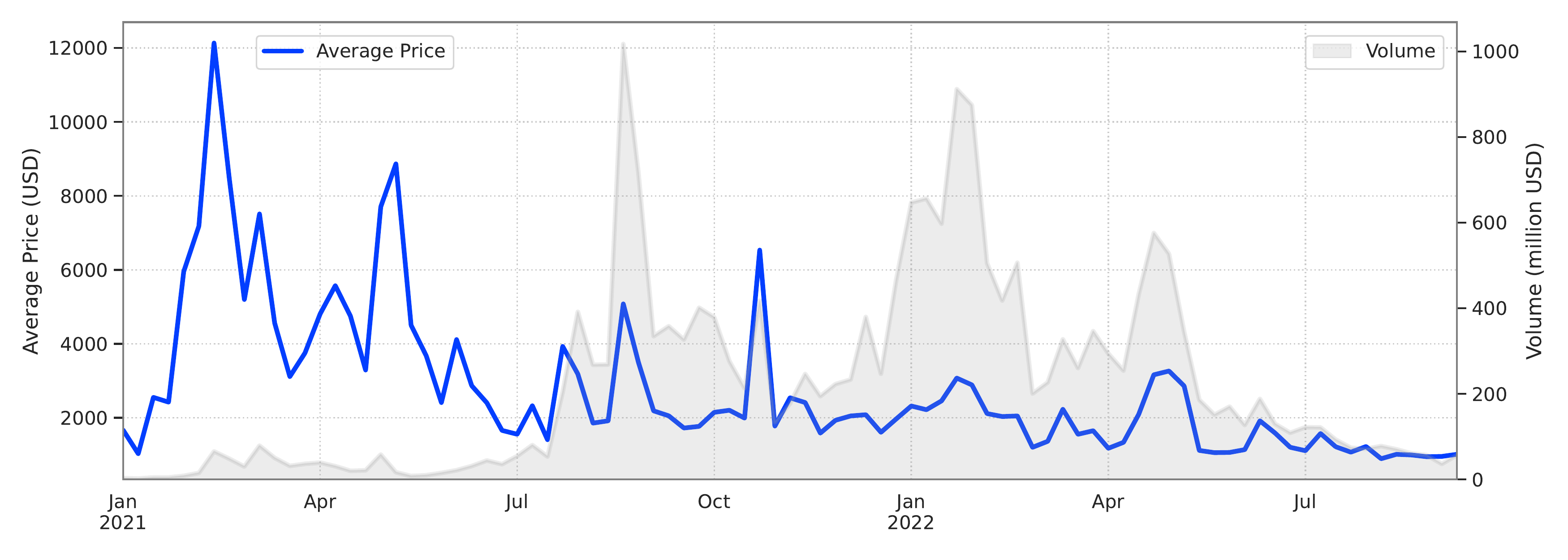}{0.95}{
        	The figure presents the weekly time series of average transaction prices (in USD) and total trading volume (in million USD) based on our data set, comprising the 1,000 most traded NFT collections on OpenSea from January 2021 to September 2022.
        	This sample covers around 5\% of the total dollar volume on NFTs across all exchanges and all blockchains over the same period.
        }
        %

	\section{Price Run-Ups and Bubbles}\label{sec:bubbles}
        We now examine the evolution of prices for our sample of NFT collections to identify price run-ups and subsequent crashes. We then run cross-sectional regressions on the identified events to assess the ex-ante predictability of crashes.
        
        \subsection{Price Run-up Events}\label{subsec:events}
		The approach used to identify price run-up events in our study is based on the methodology proposed by \cite{greenwood2019bubbles}. However, we have adapted their approach to suit our intraday data that cover the NFT market and to neatly link individual investors' behavior to high-frequency abnormal price movements that can be considered short-term bubbles, similarly to the dot-com bubble literature \citep[e.g.,][]{Ofek2003,Brunnermeier2004}. Specifically, we have made two key modifications. First, we have substituted monthly returns with high-frequency (hourly) returns. Second, we have utilized a shorter window of 24 hours to detect run-ups. This ensures that the number of observations utilized to identify run-up events remains unaltered.
        More specifically, a given collection is defined to experience a \textit{price run-up} at hour $h$ if the cumulative return over the preceding $24$ hours is greater than or equal to $100\%$. 
        We consider a window of $48$ hours centered around $h$ for each run-up event, so that the event time $t=0$ corresponds to the hour in which the run-up is identified. 
        We refer to negative event times as \textit{ex-ante} and to positive event times as \textit{ex-post}. 
        We impose the condition that two run-up events for the same collection have nonoverlapping event windows, and further, we require the trading volume during the event window to sum up to at least 10 ETH.
        The procedure identifies 1,017 price run-up events, distributed over the entire sample period but more frequent in the interval between August 2021 and May 2022, as Figure \ref{fig:events_distribution} shows.
        %
        \figurecap{Events distribution}{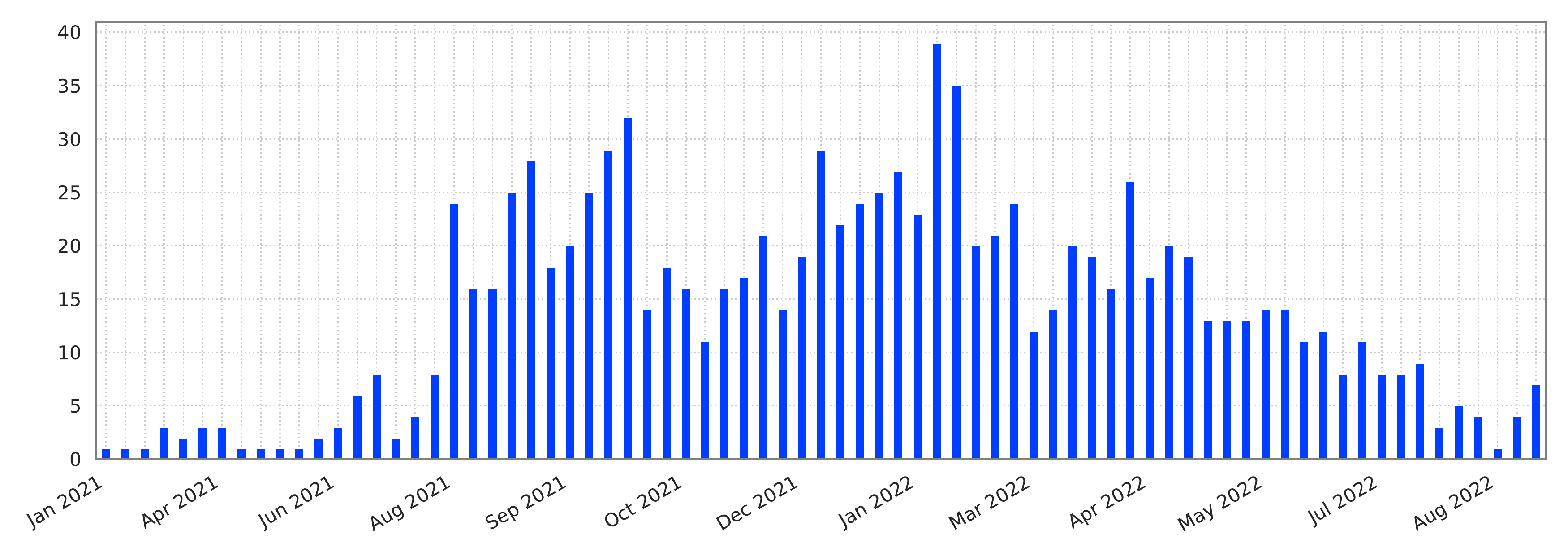}{0.95}{
        	Distribution of price run-up events over time, aggregated at a weekly frequency. Run-up events are defined as situations in which the average sell price of an NFT collection increases by more than 100\% within 24 hours.\\
        }
        Figure \ref{fig:volume_wallets} presents the distribution of event level trading volume (in USD) and the total number of active wallets per event on a logarithmic scale. The average volume per event is 1.5 million USD, with 1,300 wallets actively participating, on average.
        %
        %
        \figurecap{Events participation}{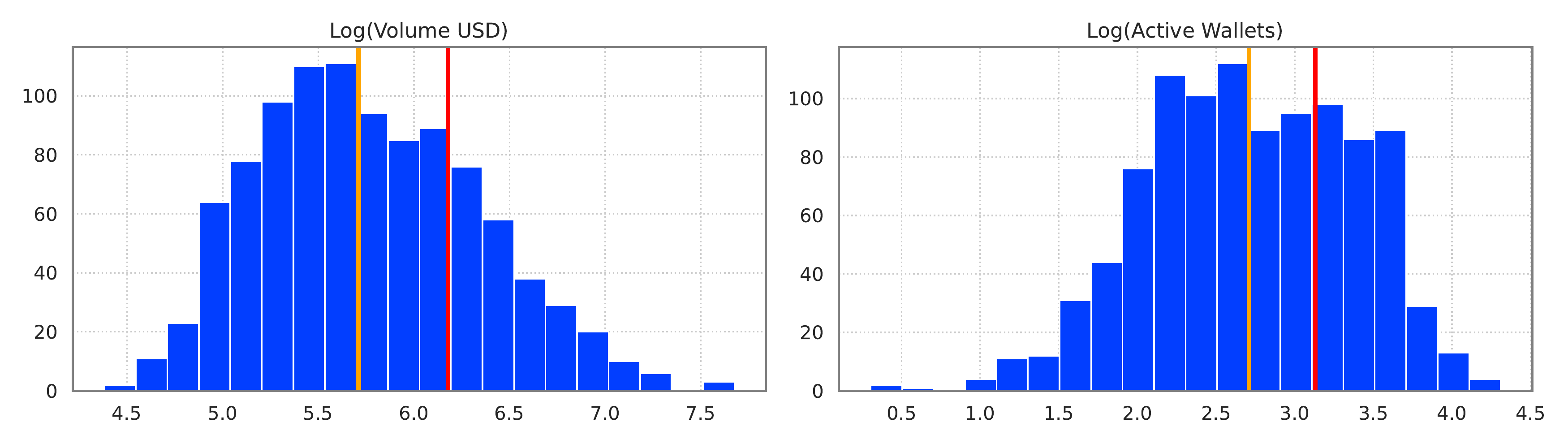}{0.95}{
			The figure shows the trading volume (in USD) and the number of active wallets per event on a logarithmic scale. The red line indicates the average, while the orange line indicates the median.            
		}
        Figure \ref{fig:events_returns.pdf} exhibits the characteristic event time pattern. On average, the event time cumulative returns are slightly positive in the ex-post period. At the same time, the median is significantly negative, implying that the distribution of ex-post returns is highly skewed. Interestingly, immediately after the run-up is identified, at $t=1$ the median is positive at around $25\%$, while it declines quickly and reaches $-50\%$ in the following four hours.
        
        \figurecap{Events returns}{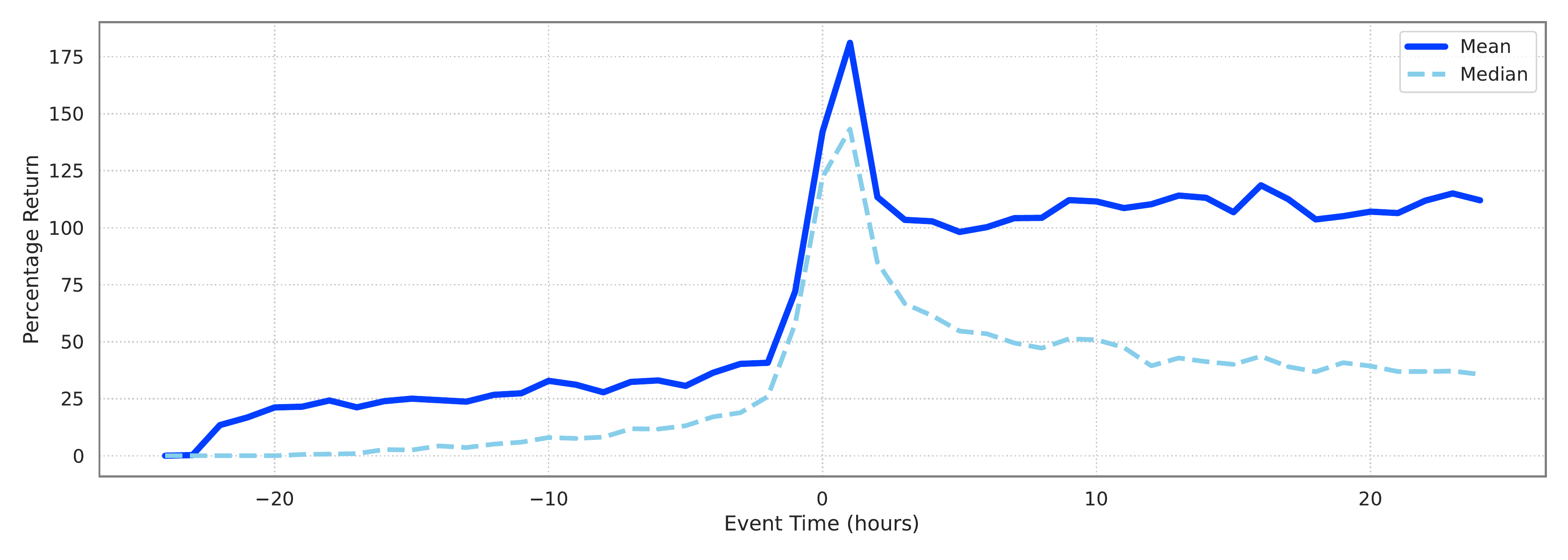}{0.95}{
            Event time average (solid, in blue) and median (dashed, in light blue) of cumulative returns for the identified events, based on changes in the average sell price of the relevant NFT collections. The event time is in hours. The central point $t=0$ corresponds to the first hour in which the cumulative return is larger than 100\%, and the window extends to the 24 hours before and after that event. The mean and median are quite different, implying that event returns are highly skewed.
        }
        
        \subsection{Crashes and Noncrashes}
        
            Next, we define an event as a \textit{crash} if the ex-post return realized during the 24 hours following the run-up identification is lower than $-40\%$.\footnote{
            	The threshold does not constitute an arbitrary choice but, rather, it is taken from \cite{greenwood2019bubbles}. Our results are robust to varying the threshold from $-20\%$ to $-80\%$.
            }
            According to this criterion, we find that about $52\%$ of the events result in a crash, ex-post. 
            This is not surprising: a high proportion of incidences was expected given the high volatility and fast-paced growth of the NFT market. 

	        \figurecap{Events distribution}{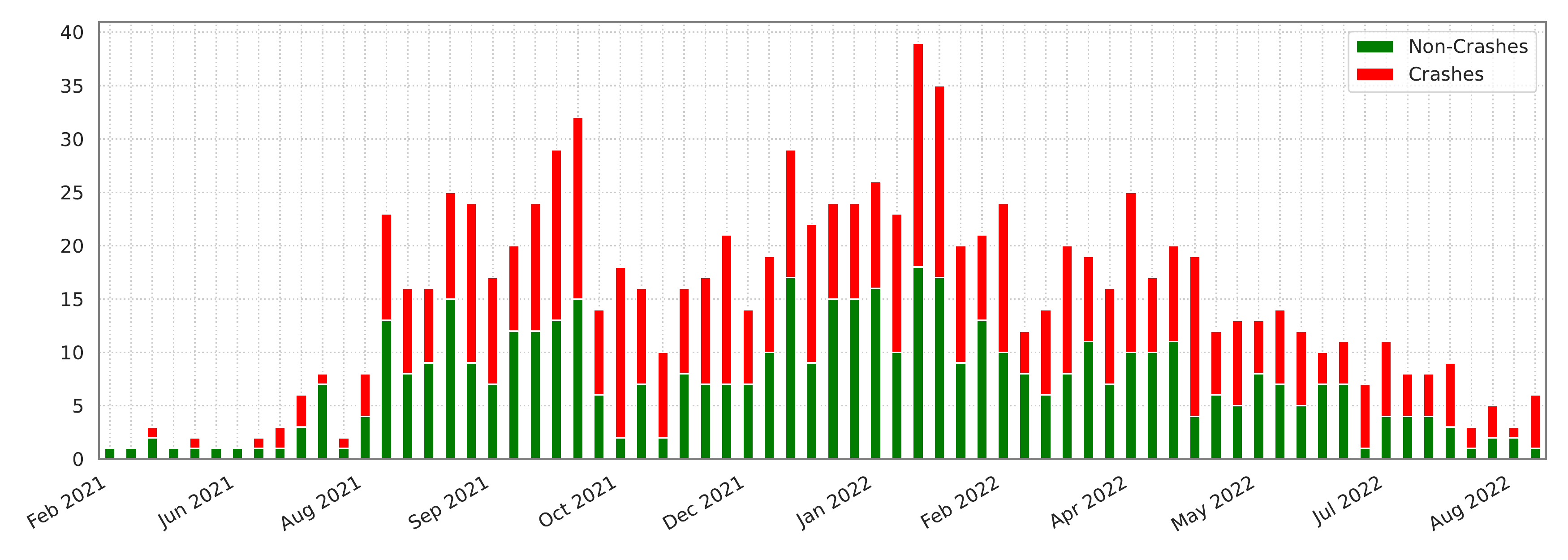}{0.95}{
	        	Distribution of price run-up events over time, aggregated at the weekly frequency and split between ex-post crashes (in red) and noncrashes (in green). Crash events are defined as those in which the ex-post return is lower than -40\% after 24 hours from the run-up identification date.
	        }

            The temporal distribution of price run-up events, classified as crashes and noncrashes, is illustrated in Figure \ref{fig:events_distribution_crashes}. The depicted clusters notwithstanding, both categories appear to be distributed uniformly over time. To further illustrate the contrasting characteristics of these two categories, Figure \ref{fig:events_returns_crashes} presents the cumulative returns of crashes and noncrashes separately. Notably, a conspicuous wedge separates the two categories. Specifically, while the appreciation in value from the ex-ante period is substantially wiped out for crashes, the noncrash events experience a significant surge in value, on average.
            %
            %
            \figurecap{Crashes}{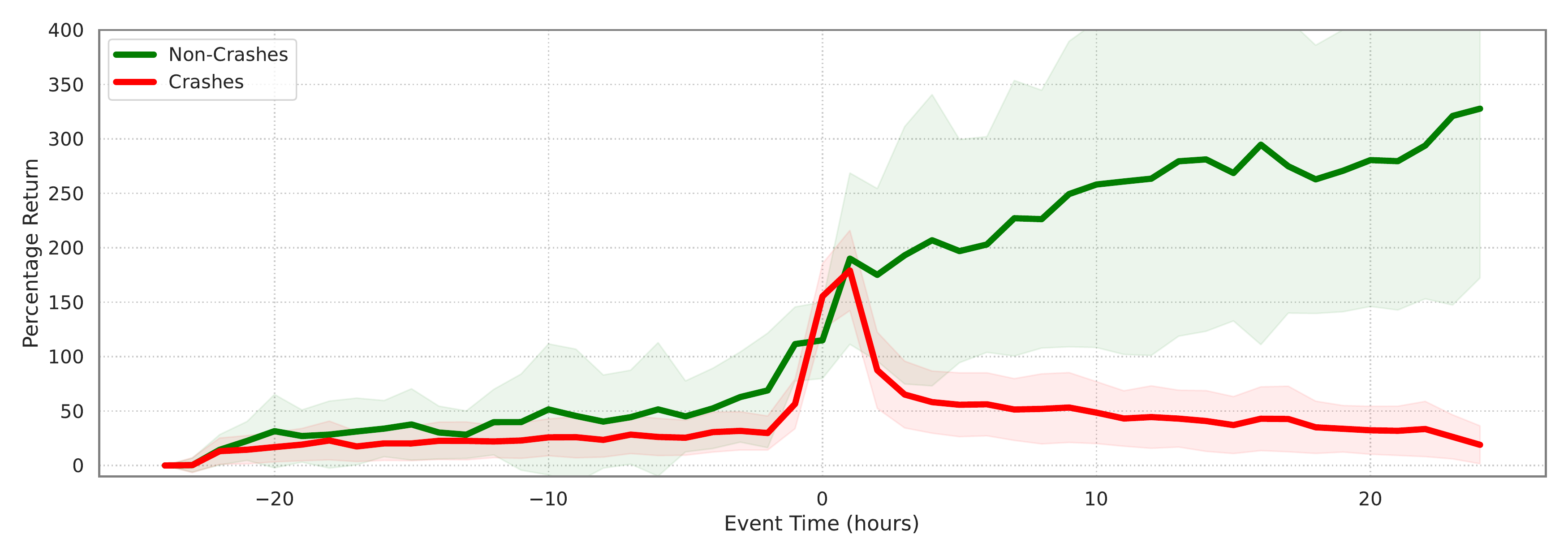}{0.95}{
				Event time (in hours) average cumulative returns for the identified events split between ex-post crashes (in red) and noncrash (in green). The central point $t=0$ corresponds to the first hour in which the cumulative return is larger than 100\%, and the window extends to the 24 hours before and after that event.
            }

            Figure \ref{fig:volume_wallets_crashes} plots the distribution of trading volume and the number of active wallets over the entire event window, split between ex-post crashes and noncrashes. 
            It is noticeable that both metrics are lower for ex-post-crash events, as it is natural to expect, given that participation and liquidity are likely to deteriorate after the bubble bursts.
            Additionally, this suggests that ex-ante liquidity levels may be related to the probability of a crash realizing ex-post. 
            We will test this hypothesis in Section \ref{sub:predicting}.
            %
			%
            \figurecap{Crashes participation}{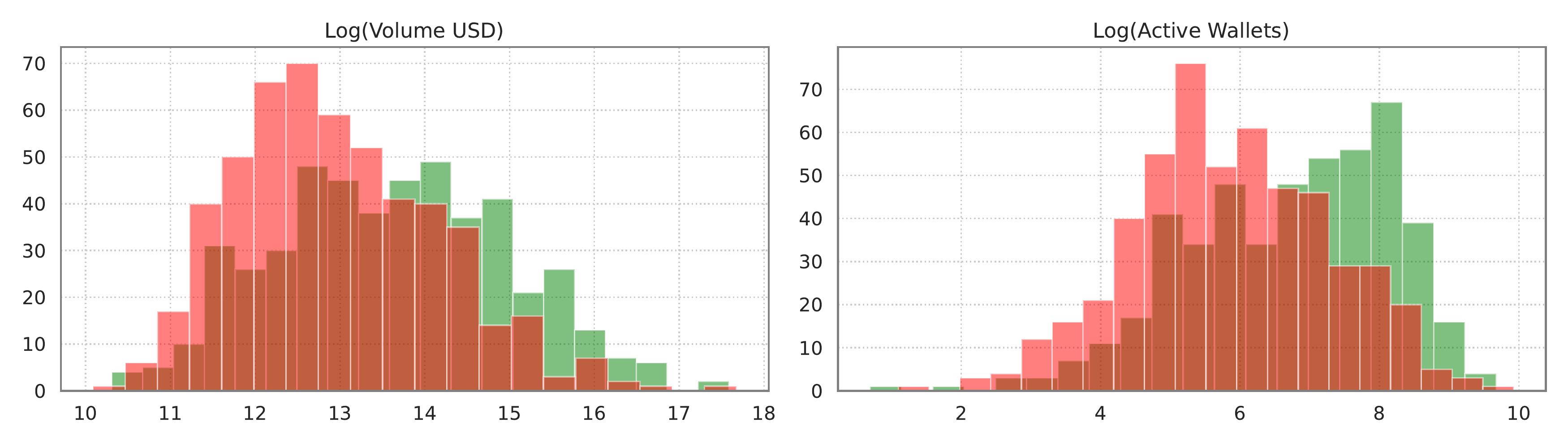}{0.95}{
				The figure shows the trading volume (in USD) and the number of active wallets per event split between ex-post crashes (in red) and non-crashes events (in green), in logarithmic scale.
            }

        \subsection{Predicting Crashes}\label{sub:predicting}
            
            In this section, we assess the ex-ante predictability of ex-post crashes. 
            More precisely, we check to see if ex-ante variables measured as of $t=0$ can help predict whether the run-up event will lead to an ex-post crash, that is, to a cumulative return from $t=1$ to $t=24$ below $-40\%$.
            We start by constructing \emph{aggregate} event-level variables that are specific to each NFT collection but do not require agent-level information. These are estimated on the ex-ante period for $t\in[-24,0]$. Building on the results of \cite{greenwood2019bubbles}, who\ identified a specific set of predictors for bubbles in the equity domain, our study centers on a specific set of variables, namely:
            \\[0.5em]
        	\textit{Volatility:} The standard deviation of hourly returns.
        	\\
        	\textit{Turnover:} The number of sales divided by the circulating supply.
        	\\
        	\textit{Age:} The number of hours since the collection launch.
        	\\
        	\textit{Acceleration:} The difference between the cumulative return realized on the interval $[-24,0]$ and the one realized on $[-24,-12]$.
			\\[1em]
            We use these variables as predictors in an event-level cross-sectional regression where the dependent variable is a dummy indicating an ex-post crash.

            The regression estimates are reported in Table \ref{tab:fama_reg}, unveiling a significant degree of crash predictability. As for equities \citep{greenwood2019bubbles}, a crash is significantly more likely to happen when the return volatility and acceleration are higher in the ex-ante period.\footnote{
            	Ex-ante volatility could be mechanically correlated with ex-post crashes since higher volatility increases the chance of realizing $\pm 40\%$ price movement. To rule out this possibility, we rerun the regression using as a dependent variable a dummy indicating ex-post returns higher than $40\%$. We find a significantly negative sign on volatility, suggesting that the potential mechanical effect, if anything, is of second-order importance.
            }
            On the contrary, the probability is negatively correlated with the level of ex-ante turnover. In a context such as NFTs, in which it is not possible to sell short, this result squares with the idea that a higher turnover indicates a lower concentration of optimistic and overconfident investors, so as to decrease the risk of a speculative bubble \citep{Scheinkman2003}. Moreover, high turnover can also predict lower liquidity risk, thus marking a smoother price adjustment process  \citep{Nneji2015}.

            In the last specification, where all variables are used as regressors, the coefficient of determination is greater than 21\%, unveiling an economically significant degree of crash predictability. 
            We will expand on this finding in Section \ref{sec:agents}, by including some additional agent-level predictors.
            %
	\begin{table}[t!]
		\centering
		\resizebox{0.94\textwidth}{!}{\begin{tabular}{lccccc}
\toprule
              & (1)     & (2)      & (3)     & (4)     & (5)      \\
Dep. Variable & $\;\quad\,$Crash$\;\quad\,$     & $\;\quad\,$Crash$\;\quad\,$      & $\;\quad\,$Crash$\;\quad\,$     & $\;\quad\,$Crash$\;\quad\,$     & $\;\quad\,$Crash$\;\quad\,$      \\
\midrule
              &           &            &           &           &            \\
Volatility    & 0.4401*** &            &           &           & 0.2714***   \\
              & (7.2525)  &            &           &           & (4.5709)    \\[1.0em]
Turnover      &           & -0.3768*** &           &           & -0.3199***  \\
              &           & (-5.5371)  &           &           & (-5.5626)   \\[1.0em]
Age           &           &            & 0.0001*** &           & 0.0001***   \\
              &           &            & (4.0524)  &           & (3.1493)    \\[1.0em]
Acceleration  &           &            &           & 0.1668*** & 0.1594***   \\
              &           &            &           & (11.344)  & (10.557)    \\[1.0em]
Intercept     & 0.2582*** & 0.5431***  & 0.4751*** & 0.3335*** & 0.1669***   \\
              & (6.6416)  & (30.182)   & (24.898)  & (15.626)  & (3.9723)    \\[0.1em]
              &           &            &           &           &            \\
Observations & 1,017 & 1,017 & 1,017 & 1,017 & 1,017 \\
R-squared             & 0.0468 & 0.0192 & 0.0217 & 0.1601 & 0.2110 \\
\bottomrule
\end{tabular}}
		\captionsetup{justification=justified, singlelinecheck=on, font=footnotesize}
		\caption{\textbf{Crash Predictability.} 
            	The table reports results from cross-sectional linear regressions based on the identified run-up events.
				The dependent variable is a dummy indicating an ex-post crash, that is, a cumulative return from $t=1$ to $t=24$ below -40\%.
				The dummy is regressed onto a set of market-level variables measured as of the price run-up identification, that is, at $t=0$, including:
				(i) \emph{Volatility}, the hourly returns standard deviation estimated between $t=-24$ and $t=0$;
				(ii) \emph{Turnover}, the average turnover ratio, that is, the number of hourly sales transactions divided by the outstanding supply, estimated between $t=-24$ and $t=0$;
				(iii) \emph{Age}, the time from the creation of the NFT collection in hours, measured as $t=0$; and
				(iv) \emph{Acceleration}, the difference between the cumulative return realized in $\left[-12,0\right]$ and the cumulative return realized in $\left[-24,0\right]$.
				\tstats
            }\label{tab:fama_reg}
		\vspace{0.25em}
	\end{table}

            These results are robust to estimating a logistic regression model with heteroskedasticity-robust standard errors. Moreover, similar results, but with opposite signs, are obtained using an alternative specification in which the cumulative return in [0,24] is used as the dependent variable. 
            The \textit{R}-squared coefficient is even larger in this alternative specification, reaching 36\%.
            Finally, the results are robust to alternative definitions of ex-post returns, adjusting by an equally weighted return index or the NFT index proposed by \cite{borri2022economics}.
            In nonreported experiments, we run alternative specifications featuring additional regressors suggested by the literature.
            For instance, we found the skewness of ex-ante returns to be a positive predictor of crashes, but its statistical significance evaporates once we control for \emph{Acceleration}. 

            Overall, our results are consistent with the empirical evidence on U.S. and international industry portfolios \citep{greenwood2019bubbles} and theories of heterogeneous beliefs with shortsale constraints, which predict that volatility and price acceleration during price run-up events are positively related to the probability of a crash. 
            The empirical findings of this study suggest that the emergence of speculative bubbles in conventional financial securities and novel digital securities, such as NFTs, adheres to similar dynamics. This suggests that the NFT market is not inherently more irrational than traditional financial markets, albeit learning about these innovative digital assets and their underlying technology may play a crucial role in this context \citep{Pastor2009}.

        \subsection{Market Factors}
        	We now test whether run-ups and bubbles are driven by common factors, for example, by returns on a broad index of the NFT market.
        	First, we apply principal component analysis to the panels of daily and weekly returns for our sample of collections. We find that the first five principal components explain less than 1\% of the variation in returns.
        	Second, we construct a market return factor by computing the daily average transaction prices across all collections of our sample and taking their percentage changes. 
        	We then run time-series regressions of collections' daily returns on the market factor.
			We find that the absolute value of the estimated market betas is, on average, 0.07 (median 0.05), and the average \textit{R}-squared coefficient is 0.60\% (median 0.05\%).
			Similar results are obtained by using returns at the hourly or weekly frequency. 
			More interesting results are obtained using as a market factor the NFT index proposed by \cite{borri2022economics}, with the absolute value of estimated market betas averaging 0.41 (median 0.19), and slightly better \textit{R}-squared averaging 3.34\% (median 0.40\%). 
        	We conclude that, consistent with \cite{borri2022economics}, returns of NFT collections are highly idiosyncratic and poorly explained by the market index alone.

        	Next, to test the hypothesis that bubbles are clustered around specific time periods and autocorrelated, we run a regression similar to that reported in Table \ref{tab:fama_reg}, but including as additional regressors a variable counting the number of run-up events recorded in the previous 1 to 10 days, and another variable measuring the likelihood of crashes for those previous events.
            In unreported regression estimates, we found that both variables are not significant predictors of crashes. 
            This is in line with the pattern displayed in Figure \ref{fig:events_distribution_crashes}, showing that bubbles are evenly distributed at a weekly frequency.
            
            All in all, we conclude that bubbles are not clustered in specific time periods and that the market factor is not the main driver of price run-ups and their continuation.
            The fact that \emph{outer} variables proxying for market conditions cannot predict crashes highlights the idiosyncratic nature of the run-up events in our sample.
            Motivated by this observation, we will take an \emph{inner} look at the events in Section \ref{sec:agents}, testing the prediction power of event-specific, agent-level variables.

        \subsection{Crashes and Liquidity Dry-Ups}
        	In this section, we evaluate the role of liquidity in speculative bubbles. Specifically, we explore two hypotheses: first, that liquidity decreases during bubble bursts, in line with the narrative about speculative bubbles \citep{Kindleberger1978}. Second, limited liquidity increases the likelihood of a speculative bubble bursting, as predicted by models with rational and financially constrained agents demanding and supplying liquidity \citep{Farhi2011}. To do so, we regress different measures of liquidity and trading activity in the ex-post period on the crash indicator dummy.
        	The measures include: (i) the average turnover ratio in the ex-post window; (ii) the \cite{Amihud2002illiquidity} illiquidity ratio, defined as the ratio between the absolute cumulative return from $t=1$ to $t=24$ and the volume traded over the same period; and (iii) the hourly volatility estimated in the ex-post window.
%
	\begin{table}[t!]
		\centering
		\resizebox{0.94\textwidth}{!}{\begin{tabular}{lccccccccc}
\toprule
 & (1) & (2) & (3) & (4) & (5) & (6) & (7) & (8) & (9) \\
Dep. Variable         & Turnover & Amihud & Volatility & Turnover & Amihud & Volatility & Turnover & Amihud & Volatility\\
\midrule
  &  &  &  &  &  &  &  &  &  \\
Crash Dummy   & -0.03***        & 0.0***        & 0.11***         &                 &               &                   &                 &               &              \\
              & (-6.04)         & (3.23)        & (7.09)          &                 &               &                   &                 &               &              \\[1.0em]
Unique Owners &                 &               &                   & 0.01***         & -0.0***       & -0.02***          &                 &               &            \\
              &                 &               &                   & (5.58)          & (-4.93)       & (-6.43)         &                 &               &              \\[1.0em]
Wash Trading  &                 &               &                   &                 &               &                   & -0.01***      & 0.01        & 0.02***         \\
              &                 &               &                   &                 &               &                   & (-8.63)       & (1.18)      & (2.85)         \\[1.0em]
Intercept     & 0.04***         & 0.01***       & 0.48***           & 0.02***         & 0.01***       & 0.56***           & 0.04***         & 0.01***       & 0.51***    \\
              & (9.53)          & (5.10)        & (35.57)           & (8.51)          & (12.14)       & (50.46)           & (10.89)         & (8.25)        & (40.57)    \\
&  &  &  &  &  &  &  &  &  \\
Observations & 1,017 & 1,017 & 1,017 & 1,017 & 1,017 & 1,017 & 1,017 & 1,017 & 1,017 \\[0.1em]
R-squared             & 0.04 & 0.01 & 0.05 & 0.12 & 0.01 & 0.04 & 0.06 & 0.00 & 0.01 \\[0.25em]
\bottomrule
\end{tabular}}
		\captionsetup{justification=justified, singlelinecheck=on, font=footnotesize}
		\caption{\textbf{Liquidity Dry-Up.} 
        	    The table reports results from cross-sectional linear regressions based on the identified run-up events.
				In specifications (1), (4), and (7), the dependent variable is the turnover ratio, that is, the number of hourly sales transactions divided by the outstanding supply, estimated in the ex-post window \postwin.
				In specifications (2), (5), and (8), the dependent variable is the Amihud ratio, defined as the ratio between the absolute cumulative return from $t=1$ to $t=24$ and the volume traded over the same period.
				In specifications (3), (6), and (9), the dependent variable is the hourly volatility estimated as the standard deviation of hourly returns in the ex-post window \postwin.
				In specifications (1) to (3), the dependent variables are regressed onto the dummy indicating an ex-post crash, that is, an ex-post cumulative return lower than $-40\%$.
				In specifications (4) to (6), the dependent variables are regressed onto the change in the ratio of unique owners and the supply of NFT tokens from $t=-24$ to $t=0$.
				In specifications (7) to (9), the dependent variables are regressed onto the natural logarithm of the amount of wash-trading volume identified for the collection before $t=0$.
				\tstats
        	}\label{tab:reg_ex_post_liquidity}
		\vspace{0.25em}
	\end{table}

        	Table \ref{tab:reg_ex_post_liquidity} presents the regression results of the first three specifications (1) to (3), indicating significantly lower liquidity in the ex-post period for crash events, which is consistent with the liquidity dry-up predicted by the asset bubble narrative.
			Next, to explore the predictability of ex-post liquidity by ex-ante variables, we incorporate an inverse measure of the change in ownership concentration during the price run-up --- referred to as \emph{Unique Owners} --- as the explanatory variable in specifications (4) to (6). 
			This variable is extensively defined and analyzed in Section \ref{sub:ownership}, and a higher number of unique owners is interpreted as a larger and more diverse trading community that facilitates the resale of financial securities, reduces search costs, and increases risk sharing. The regression results suggest that a decreasing number of owners ex-ante positively predicts ex-post illiquidity.
			Finally, specifications (7) to (9) employ the ex-ante wash-trading volume as a regressor, as defined and estimated in Section \ref{sub:wash}. 
			The estimated coefficients indicate that higher ex-ante wash trading is linked to aggravated ex-post illiquidity. These findings corroborate the notion that the fictitious volume generated by wash trades during the run-up phase may culminate in more illiquid trading during the correction phase, underscoring the importance of heterogeneous and genuine investors in invigorating the market.

	\section{Agent-level Analysis}\label{sec:agents}
		The objective of this section is to develop a set of metrics for analyzing individual investor behavior in the NFT market, and to explore their predictive power for identifying crashes. 
		To this end, we construct three agent-level variables based on economic arguments. 
		The first variable identifies a group of sophisticated investors, sorting by the realized profits during run-up events. We validate their superior abilities in the wider decentralized finance (DeFi) space by analyzing their blockchain transactions, and examine the relation between their involvement in run-ups and the occurrence of crashes.
		The second variable is a measure of ownership concentration, which is determined by counting the unique owners of a collection. We test whether this metric denoting investor heterogeneity is related to crash probability and ex-post illiquidity. 
		The third variable involves identifying suspicious transactions that are likely used to artificially inflate the trading volume of a collection (wash trading) and testing whether the degree of this phenomenon is a predictor of ex-post crashes and illiquidity.\footnote{
			It is important to acknowledge that individual investors may possess multiple wallets, which could introduce errors into the agent-level variables constructed using data at the wallet-address level. However, we contend that such additional noise is likely to hinder the identification of statistically significant predictors of crashes.
		}

		\subsection{Sophisticated Agents}\label{sub:sophisticated}

			The motivation underlying this analysis stems from the narrative surrounding speculative bubbles, as documented in the literature \citep[e.g.,][]{Kindleberger1978,Brunnermeier2016}. This narrative posits that market participants can be broadly classified into two groups: speculators and sophisticated investors. While the former are primarily interested in the resale price, the latter possess the ability to link the market value of a financial security to its fundamental value. However, in the case of NFTs, identifying the fundamental value is challenging due to the absence of cash flow streams, the unique characteristics of these digital assets, and the innovative nature of the underlying technology. When speculators become the dominant force in the market, the likelihood of a bubble occurring increases.
			The existing literature offers various aspects that differentiate economic agents, including but not limited to asymmetric information \citep[e.g.,][]{ALLEN1993206,10.1257/jep.21.2.109}, arbitrage \citep[e.g.,][]{https://doi.org/10.1111/j.1540-6261.1997.tb03807.x,abreu2003bubbles}, heterogeneous beliefs \citep[e.g.,][]{Scheinkman2003}, the ability to time the market \citep[e.g.,][]{Brunnermeier2004}, and risk premia combined with learning about new technologies \citep[e.g.,][]{Pastor2009}. Our underlying assumption is that these factors may give rise to heterogeneity among investors, which, in turn, can affect their performance consistently over time.
	  
	    	As a preliminary step, we aim to investigate whether certain investors consistently outperform others, that is, whether there exists a group of sophisticated investors who possess the ability to generate gains by riding successive price run-ups while avoiding price corrections. This would serve as a first indication of their sophistication.
			To this end, we compute the percentage profits obtained by individual investors during the entire run-up event window of $[-24,24]$. We find that the profitability of investors is strongly autocorrelated across different run-up events, with a significant autoregressive coefficient of 0.21 (\textit{t}-stat = 27.59). These results imply that some investors consistently excel at riding price run-ups and avoiding crashes, while others consistently perform poorly.
			Motivated by these findings, we define a group of \textit{sophisticated} investors based on their past performance. Specifically, for each event $e$ and investor $a$, we consider the investor's profitability in the previous five events in which they actively participated. We require sophisticated investors to have participated in at least five run-up events and to have realized an average profit of more than 25\% on their invested capital during the previous five events. Profitability is calculated over the entire event window, from $t=-24$ to $t=24$. If an NFT is not sold before $t=24$, its profit is based on the average transaction price at $t=24$.
			We define the first agent-level variable as the percentage of sophisticated investors actively participating in the ex-ante period of a run-up event, relative to the total number of active wallets during the same period.

        	Before moving to the definition of the remaining two agent-level variables, the next subsections provide more information about the identified sophisticated agents, focusing on their skills in the broader DeFi ecosystem and their ability to properly time NFT bubbles.
			%

        	\subsubsection{Characteristics of Sophisticated Agents}
    			We now conduct a detailed analysis of the behavior of sophisticated agents, so as to conjecture whether they have superior abilities even outside of the NFT scope. As discussed earlier, factors that might distinguish them are (a) superior information extracted from trading or learned by observing the price discovery process or (b) honed skills in conducting their trading activities. 
	    		To this end, we have augmented our existing data set by procuring a full record of all transactions on the Ethereum blockchain conducted by each wallet in our sample from the platform \href{https://www.etherscan.io}{Etherscan}. These transactions encompass a wide range of activities, including not only the straightforward transfer of ether or ERC-20 tokens, but also interactions with diverse smart contracts, even those that are not directly related to NFTs.
			
	            Drawing on our extensive data set, we proceed to investigate two key determinants that are potentially related to the level of sophistication exhibited by market agents. We explore the impact of two factors, namely, the frequency and value of transactions and the agents' proficiency in certain trading activities, on the performance of sophisticated traders in NFT markets.
	            First, we contend that increased levels of trading activity enable sophisticated agents to extract a greater volume of \textit{information} from the market, thereby enhancing their capacity to acquire a more nuanced understanding of fundamental value compared to market price evolution during the price discovery process. In this vein, we scrutinize the extent to which the superior performance of sophisticated agents correlates with more frequent and valuable transactions that are primarily attributable to informed traders, according to microstructure theories of market dynamics \citep[e.g.,][]{EASLEY198769}.
	            Moreover, we surmise that the extended presence of sophisticated agents in the market provides a distinct advantage, especially in the context of nascent assets like NFTs, whose fundamental value remains obscure.
	            Second, we posit that sophisticated traders exhibit superior \textit{ability} in at least three domains, namely, trading with precise timing to maximize gains, leveraging their positions to exploit profit opportunities and access arbitrage capital, and engaging in cross-market trading.

	        	Table \ref{tab:sophi_diff} presents an overview of the average characteristics of agents who are identified as \emph{sophisticated} in at least one event, taking into account NFT-related metrics, such as profits, trading volume, holding period, and the number of NFT transactions, in conjunction with supplementary metrics derived from the blockchain data. Specifically, we construct these metrics by counting interactions with a set of DeFi-related smart contracts, including those unrelated to NFTs, over the entire lifespan of each wallet account and measured as of December 2022.

				The resulting metrics comprise the number and total value of transactions conducted in ETH, the wallet's age measured in days, as well as swaps and liquidity provisions on decentralized cryptocurrency exchanges (DEX) such as Uniswap or Curve, and lending or borrowing operations on decentralized lending platforms such as AAVE or Compound. We note that a similar set of results can be obtained by focusing only on blockchain transactions conducted before the start of our sample period, which is January 2021.	        	
				%
	\begin{table}[t!]
		\centering
		\resizebox{0.94\textwidth}{!}{\begin{tabular}{lrrrr}
\toprule
                           & Sophisticated & Others & Difference & t-stat    \\
\midrule
\color{electric_blue}{\textbf{NFT Market}}\color{black} \\
Number of Trades           $\qquad\qquad$& 22.41         & 3.05   & 19.36      & 136.26*** \\
Traded Volume (ETH)        $\qquad\qquad$& 8.49          & 1.43   & 7.06       & 98.37***  \\
Percentage Profits         $\qquad\qquad$& 94.92         & 69.41  & 25.51      & 44.59***  \\
Holding Period (hours)     $\qquad\qquad$& 7.76          & 5.72   & 2.04       & 62.25***  \\[1em]

\color{electric_blue}\textbf{Blockchain}\color{black} \\
Number of Transactions     $\qquad\qquad$& 1,088.48      & 273.17 & 815.30     & 123.40*** \\
Transacted Value           $\qquad\qquad$& 631.61        & 220.70 & 410.91     & 13.83***  \\
Age (days)                 $\qquad\qquad$& 519.13        & 429.89 & 89.24      & 50.24***  \\
DEX Swaps                  $\qquad\qquad$& 75.24         & 24.40  & 50.83      & 34.59***  \\
DEX Liquidity              $\qquad\qquad$& 3.58          & 1.25   & 2.32       & 19.64***  \\
DeFi Lending               $\qquad\qquad$& 0.30          & 0.08   & 0.22       & 6.17***   \\
DeFi Borrowing             $\qquad\qquad$& 0.79          & 0.27   & 0.52       & 6.99***   \\
\bottomrule
\end{tabular}}
		\captionsetup{justification=justified, singlelinecheck=on, font=footnotesize}
		\caption{\textbf{Sophisticated Agents.} 
	        		The table displays the average characteristics of the wallet accounts participating in our run-up events, split between sophisticated and nonsophisticated investors, defined in Section \ref{sub:sophisticated}.
	        		The first set of characteristics (rows 1 to 4) is constructed using data from our run-up events only and includes the following:
	        		the number of trades,
	        		the generated trading volume, 
	        		the hourly percentage profits accrued by trading NFTs, and
	        		the holding period for single NFTs.  
	        		The second set of characteristics (rows 5 to 11) is constructed using blockchain data comprising all transactions performed by those wallet accounts on the Ethereum blockchain, including those unrelated to NFTs, across their entire existence, as of December 2022.
	        		This set includes:
	        		the total number of transactions (including simple ETH transfers and interactions with smart contracts),
	        		the total ETH value attached to those transactions,
	        		the number of days that elapsed since the first transaction performed by the wallet,
	        		the number of swap transactions in decentralized exchanges (DEX; specifically, Uniswap v2, Uniswap v3, Sushiswap v2, Metamask Swap, Ox Exchange, OneInch v2, OneInch v3, OneInch v4, Balancer, and Curve),
	        		the number of liquidity deposits and withdrawals in the same list of exchanges, and the number of lending and borrowing operations in decentralized lending platforms (AAVE v1, AAVE v2, and Compound).
	        		Qualitatively similar differences are obtained by using only transactions performed before the beginning of our samples, as of January 2021.
	        		The \textit{t}-stats of the differences are reported in the last column. Asterisks denote significance levels (***$ =1\%$, **$ =5\%$, *$ =10\%$).
	        	}\label{tab:sophi_diff}
		\vspace{0.25em}
	\end{table}

	         	Table \ref{tab:sophi_diff} presents evidence that agents who outperform, as defined above based on their realized profits during price run-up events, display greater market presence both in the time series and across different DeFi applications, as well as higher levels of financial leverage.
				In particular, these agents exhibit greater sophistication, as evidenced by the fact that they created their wallets earlier and have conducted more transactions on the blockchain in terms of both size and frequency. These findings are consistent with previous empirical evidence on the formation of bubbles in traditional financial securities \citep{GREENWOOD2009239} and laboratory experiments \citep{10.2307/1911361}.
				Furthermore, outperforming agents interact more frequently with DeFi protocols, such as decentralized exchanges, which suggests cross-market ability and information extraction. Most importantly, they engage in more decentralized lending platform activities, which indicates a more proactive use of leverage and arbitrage capital.
				Taken together, these results suggest that our definition of sophisticated investors, based on NFT trading performance, captures individual retail investors who are more experienced and sophisticated agents in the broader DeFi space.

       		\subsubsection{Timing the Bubble}\label{sub:timing}
	        	Another trait usually associated with sophisticated traders is the ability to \emph{time the bubble}.
				The objective of this section is to test this hypothesis by examining whether they skillfully time the bubble during ex-post crash events, constructing an agent-level measure of market timing, $TS$, standing for \emph{Timing Score}.
				To define the measure, for each ex-post crash, we first identify the peak of the bubble, that is, the event time at which the average sale price reaches its maximum in the ex-post window following the price run-up identification.
				Denote the peak time as $t^*$, expressed in event time units (at the hourly frequency).
				We then define the distance to the peak as a translation of the event time as $d=t-t^*$.
				The timing scores for buy and sell orders are two piece-wise linear functions of $d$, with a kink at $d=0$, defined on the window $d\in [-24,24]$ centered around the peak time.
				Let us denote these two functions by $TS_{buy}(d)$ and $TS_{sell}(d)$, respectively, and define them as
				\begin{equation}
					\begin{cases}
					    TS_{buy}(d) = -d -12 & \text{for } d \leq 0
					    \\
					    TS_{buy}(d) =  d -12 & \text{for } d >    0
					\end{cases}
				\end{equation}
				and $TS_{sell}(d) = -TS_{buy}(d)$. 
				In other words, the timing score for buy orders decreases linearly while moving closer to the peak, and it increases linearly after the peak. This is in line with the intuition that it is better to buy early in order to fully profit from the price run-up.
				The opposite holds for a sell order, where perfect timing involves selling as close as possible to the peak.
				We compute the timing score for an investor participating in a given event as
				\begin{equation}
					TS = \sum_{d=-24}^{24} B(d) \cdot TS_{buy}(d) + S(d) \cdot TS_{sell}(d), 	
				\end{equation}
				where $B(d)$ and $S(d)$ count the number of buy and sell trades performed by the investor at time $d$, respectively.
				The specified definition of the timing score implies that any investor who buys and sells an NFT within the same hour will not receive a positive contribution to their timing score. This restriction is necessary to ensure the accuracy and usefulness of the metric.

				We compute the timing score $TS$ for all investor-event pairs, restricting to crash events.
				Additionally, we compute the \emph{buy} and \emph{sell} legs separately.
				Next, we compute percentile rankings by event for the obtained metrics to properly compare investors' performance within each event.
				We then run a panel regression of the $TS$ metrics on a dummy indicating sophisticated investors.
				Results are reported in Table \ref{tab:reg_timing} and show that sophisticated investors achieve better timing scores across all specifications.
				%
	\begin{table}[t!]
		\centering
		\resizebox{0.94\textwidth}{!}{\begin{tabular}{lcccccc}
\toprule
 & (1) & (2) & (3) & (4) & (5) & (6) \\
Dep. Variable         & Timing & Buy Timing & Sell Timing & Timing & Buy Timing & Sell Timing \\
\midrule
  &  &  &  &  &  &  \\
Sophisticated (time-varying)  & 0.0209***    & 0.0100***        & 0.0107***        &              &                  &                  \\
                              & (12.791)     & (6.0698)         & (14.365)         &              &                  &                  \\[1.0em]
Sophisticated (constant)      &              &                  &                  & 0.0204***    & 0.0080***        & 0.0118***        \\
                              &              &                  &                  & (16.356)     & (6.3915)         & (22.123)         \\[1.0em]
Intercept                     & 0.4971***    & 0.4992***        & 0.4991***        & 0.4930***    & 0.4979***        & 0.4965***        \\
                              & (839.90)     & (846.56)         & (2077.1)         & (766.76)     & (778.76)         & (1998.2)         \\[1.0em]
  &  &  &  &  &  &  \\
Observations & 214,855 & 214,855 & 214,855 & 214,855 & 214,855 & 214,855 \\
R-squared             & 0.0011 & 0.0003 & 0.0015 & 0.0016 & 0.0003 & 0.0028 \\
\bottomrule
\end{tabular}
}
		\captionsetup{justification=justified, singlelinecheck=on, font=footnotesize}
		\caption{\textbf{Market Timing.} 
	            	The table reports results from linear regressions at the agent-event level, restricting to the sample of crash events.
					The dependent variables are proxies for the investor's ability to time the bubble, whose definition is described in the text.
					The proxies are regressed onto a dummy variable indicating investors' sophistication.
					In specifications (1) to (3), the explanatory variable is the time-varying sophistication indicator defined in \ref{sub:sophisticated}, while 
					in specifications (4) to (6) the dummy is time-invariant, that is, an agent is considered sophisticated if she is marked as such at least once in the time-varying version.
					Standard errors are clustered by event.
					\tstats
	            }\label{tab:reg_timing}
		\vspace{0.25em}
	\end{table}

				The coefficients on specifications (1) and (4) imply that sophisticated agents enjoy, on average, a higher percentile ranking of the within-event timing score by $2$ percentage points.
				We thus conclude that sophisticated investors are more skilled in timing the bubble during ex-post crash events by both buying early into the run-up phase and selling close to the peak of the bubble.

    	\subsection{Ownership Concentration}\label{sub:ownership}

    		To construct a second agent-level variable for each run-up event, we compute the \emph{fraction of unique owners} of the relevant collection, at an hourly frequency, defined as the ratio between the number of unique wallet addresses holding at least an NFT of the collection and the total number of NFTs available.
    		This metric, displayed on OpenSea and other prominent marketplaces for NFTs, can be interpreted as an inverse measure of ownership concentration or investor heterogeneity.
    		It is expected to be a positive predictor of future performance for a number of reasons.
    		First, in the same spirit as the stock price fragility in \cite{greenwood2011stock}, a high fraction of unique owners indicates the lack of large holders that may harm the collection's value by putting a large number of NFTs on sale, potentially motivated by a liquidity shock.
    		Second, a larger and more diverse community of traders should facilitate the resale of financial securities, reduce search costs, and increases risk sharing. Thus, it positively correlates with the level of liquidity of the collection, indicating a large number of potential buyers and a high level of interest in the project.
    		Finally, it proxies for the size of the community supporting the NFT project.
    		During a price-run up, an increase in unique holders may signal that the rally comes from a genuine interest in the project.
    		On the contrary, a run-up accompanied by a decrease in unique holders signals that the price dynamic is driven by a limited number of buyers who could potentially influence or even manipulate prices in their favor, an issue we study below.
    		We thus use the change in the fraction of unique owners during the price run-up (from $t=-24$ to $t=0$) as an additional variable to predict the ex-post price evolution.

    	\subsection{Wash Trading}\label{sub:wash}

			\figurecap{Wash trading across all exchanges}{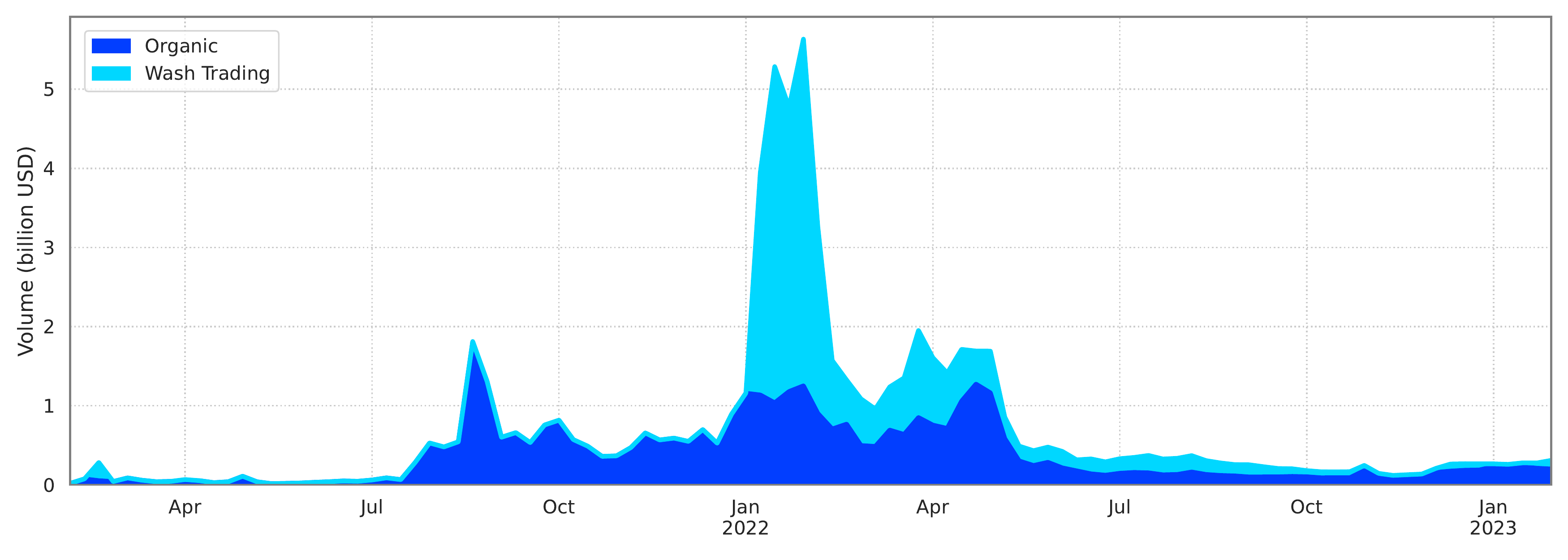}{0.95}{
	        	The figure displays the daily historical trading volume in USD across all NFT marketplaces,
	        	split between \emph{wash-trading} and organic trading volume.
	        	The former is estimated using the Hildobby method, as proposed in \cite{hildobby.2022}, which incorporates four transaction-level conditions to identify potentially suspicious trades.
	        }

	        According to the American Commodity Futures Trading Commission, wash trading is the practice of “entering into, or purporting to enter into, transactions to give the appearance that purchases and sales have been made, without incurring market risk or changing the trader's market position” \citep{CFTC}. Wash trading could be especially prevalent on cryptocurrency (unregulated) exchanges, with previous research estimating that it may account for more than $70\%$ of total trading volume on unregulated exchanges \citep{cong2021tokenomics}. 
	        %
	        In the NFT market, wash trading is usually conducted to artificially inflate prices within a collection by repeatedly trading tokens and thus increasing the total volume of the collection. 
	        These potential dynamics are relevant to our work for two reasons:
	        First, wash trading may foster artificial increases in collections' valuations, thus increasing the probability of a subsequent crash.
	        Second, neglecting such a practice could harm the identification of sophisticated investors based on agent-level profitability.
	        We therefore proceed and identify wash-trading transactions, with the two-sided purpose of (i) computing an additional agent-level predictor of crashes and (ii) testing the robustness of our results to the presence of wash trading.
	        
	\begin{table}[t!]
		\centering
		\resizebox{0.94\textwidth}{!}{\begin{tabular}{lrrrrrrr}
            & Total Volume        & Wash Volume         & Wash \% &  & Total Trades & Wash Trades & Wash \% \\
\midrule
OpenSea     & 32,797,993,136 \$ & 761,535,142 \$    & 2.32\%         &  & 27,270,308   & 181,410     & 0.67\%         \\
LooksRare   & 26,590,247,528 \$ & 26,064,869,084 \$ & 98.02\%        &  & 115,288      & 27,247      & 23.63\%        \\
X2Y2        & 4,025,756,413 \$  & 3,432,460,213 \$  & 85.26\%        &  & 587,407      & 125,241     & 21.32\%        \\
CryptoPunks & 2,495,977,352 \$  & 28,220,202 \$     & 1.13\%         &  & 22,427       & 328         & 1.46\%         \\
Blur        & 1,311,868,278 \$  & 142,010,668 \$    & 10.83\%        &  & 1,766,529    & 27,044      & 1.53\%         \\
\bottomrule
\end{tabular}}
		\captionsetup{justification=justified, singlelinecheck=on, font=footnotesize}
		\caption{\textbf{Wash Trading by Exchange.} 
	        	The table presents statistics for the top five NFT marketplaces by trading volume during the period from February 1st, 2021, to February 9th, 2023, 
	        	based on data provided by Dune Analytics.
	        	The first column reports the total traded volume in USD,
	        	while the second column reports the volume marked as \emph{wash trading} according to the Hildobby method, proposed in \cite{hildobby.2022},
	        	and the third column displays the percentage of wash-trading volume relative to the total.
	        	Columns (4) to (5) repeat the exercise using transaction count instead of transaction volume.
	        }\label{tab:wash_exchanges}
		\vspace{0.25em}
	\end{table}

	        Figure \ref{fig:NFT_wash} shows the evolution of estimated wash-trading volume over time. There is a noticeable increase in volume concentrated in early 2022, but much of the time the volume remains below \$1 billion. Table \ref{tab:wash_exchanges} shows that most of the wash trading is performed on exchanges other than OpenSea.
	        While wash-trading volume on OpenSea --- across all existing collections --- is estimated at around $2\%$ of the total, other exchanges feature much higher shares of fraudulent transactions. For instance, this applies to LooksRare and X2Y2 for which rates of $98\%$ and $85\%$, respectively, have been estimated. 
	        When the number of wash transactions is considered, the same pattern holds, with only $0.67\%$ of the transactions in OpenSea being identified as wash trades.
	        Since our data set covers only transactions performed in OpenSea, we expect a relatively low share of wash volume.
	        Nevertheless, we use two distinct approaches to estimate the magnitude of such a phenomenon quantitatively. 
	        The first is inspired by \cite{cong2021tokenomics} and provides an aggregate test for the presence of wash-trading transactions. It involves testing whether the distribution of transactions’ amounts follows patterns that are typical of such distributions. Benford's Law, frequently used to identify fraud, states that the probability of a digit from 1 to 9 appearing as the leading digit of a number adheres to the formula:
	        \begin{eqnarray}
	            \operatorname{Prob}(N \text { is the first significant digit })=\log _{10}\left(1+N^{-1}\right),
	        \end{eqnarray}
	        with $N \in\{1,2,3,4,5,6,7,8,9\}$.
	        We test Benford's Law applied to transaction prices in our sample, both in USD and ETH denominations. The leading digits' distribution in our sample shows no significant deviation from Benford's Law, with Chi-statistic tests’ \textit{p}-values approaching unity.\footnote{
				We also estimate the power-law cumulative density function exponent. Research has shown that such an exponent lies in the range from $1$ to $2$ for traditional assets and cryptocurrencies \citep{cong2021tokenomics}. In our case, the exponent value amount to approximately $2.16$, which constitutes a small deviation from the aforementioned range.
	        }
	        Based on transaction prices, Figure \ref{fig:benford.png} shows the difference between the theoretical distribution predicted by Benford's Law and the observed distribution.
	       	\figurecap{Benford's Law}{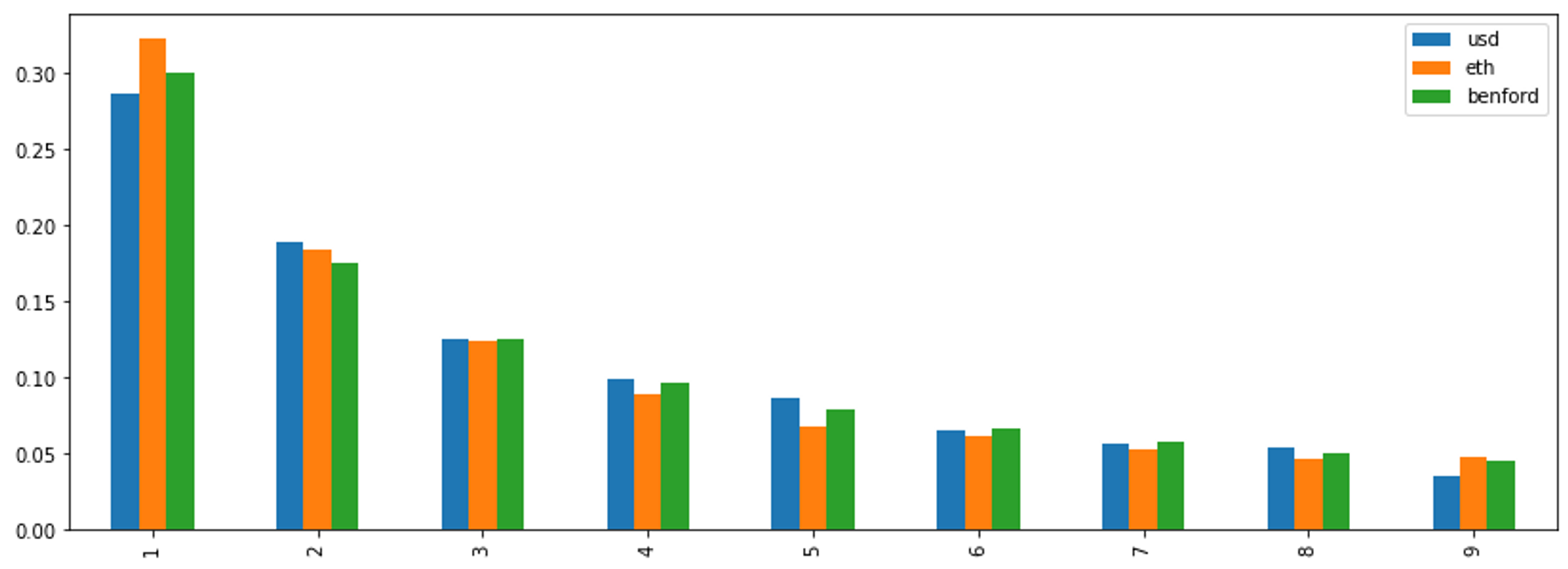}{0.95}{
	            The observed distribution of the leading digits of USD (in blue) and ETH (in orange) transaction prices and the theoretical distribution predicted by Benford's Law (in green).
	        }   
	        The second approach that we employ is based on the method proposed by \cite{hildobby.2022}, which is tailored for blockchain transactions.
			By capitalizing on the transparency of the decentralized ledger, this approach enables a transaction-level identification of wash trading.
			The methodology involves implementing four filters on our database of transactions to identify potentially dubious trading behavior.
	        Those are:
			(i)   the buyer and seller wallet addresses are identical,
			(ii)  a pair of trades with seller and buyer inverted for the same NFT,
			(iii) the same address has bought three or more times the same NFT, and (iv) the addresses that first funded the wallets of buyer and seller with ETH are identical or are one another.\footnote{
				While applying condition (iv), a list of roughly 500 addresses belonging to centralized crypto exchanges and all existing tornado cash smart contracts are excluded.
			}
			We apply this methodology restricting to our sample of collections and the time period we cover in the analysis. 
			The process results in the identification of $33,293$ transactions ($0.23\%$ of the sample), 
			corresponding to a trading volume of $63$ million \$ ($0.35\%$ of the total).
			The fraction of wash trading is insignificant for the median collection in our sample ($0.22\%$). 
			Only the \emph{Og-Crystals} collection stands out, with the estimated wash-trading volume accounting for $73.6\%$ of the total.
			As a robustness check, we also apply the methodology employed in \cite{la2022nft}, based on graph theory. 
			The transactions identified by the two methods are highly correlated, but the Hildobby method is less conservative, and the identified wash volume is more than two times larger. The methodology can thus be thought of as an upper bound for the real amount of detectable wash trading.

			We use the resulting set of wash-trading transactions to construct a variable at the event level.
			Namely, for each price run-up event, we define the \emph{Wash Volume} metric as the natural logarithm of one plus the total wash volume transacted (in ETH units) on the relevant collection before the run-up identification hour ($t=0$). 
			This variable will be used as a regressor in the cross-sectional regression of Section \ref{sub:agent_based_pred}.

		\subsection{Agent-level Crash Predictability}\label{sub:agent_based_pred}

        	We now turn our attention to the question of whether agent-level variables can serve as predictors of asset bubbles.
			To investigate this, we incorporate the three agent-level variables studied above as additional regressors in a regression specification similar to that of \ref{sub:predicting}, but with the crash indicator as the dependent variable.
			First, we add the fraction of sophisticated agents active during the event in the ex-ante window $[-24,0]$, denoted as \emph{Sophisticated}. The underlying sophistication indicator is based on the realized returns on the previous five events, as discussed earlier.
			Second, we include the change in the fraction of unique owners recorded during the ex-ante window $[-24,0]$, denoted as \emph{Unique Owners}.
			Third, we incorporate the total wash-trading volume realized for the collection before $t=0$, denoted as \emph{Wash Trading}.
        
	\begin{table}[t!]
		\centering
		\resizebox{0.94\textwidth}{!}{\begin{tabular}{lcccccc}
\toprule
              &   (1)      &   (2)      &   (3)      &   (4)          &   (5)          &   (6)         \\
Dep. Variable & Crash        & Crash          & Crash        & Ex-Post Return     & Ex-Post Return     & Ex-Post Return \\
\midrule
  &  &  &  &  &  &  \\
Sophisticated & -0.5114***   &                & -0.4155***   & 2.1993***    &              & 1.7250***   \\
              & (-3.4262)    &                & (-2.8422)    & (4.8323)     &              & (3.8923)    \\[1.0em]
Unique Owners & -0.0114*     &                & -0.0070      & 0.0585***    &              & 0.0403**    \\
              & (-1.8545)    &                & (-1.0926)    & (3.1115)     &              & (1.9849)    \\[1.0em]
Wash Trading  & 0.0770***    &                & 0.0613***    & -0.2718***   &              & -0.1988***  \\
              & (6.6193)     &                & (5.0739)     & (-7.7302)    &              & (-6.4099)   \\[1.0em]
Volatility    &              & 0.2714***      & 0.1785***    &              & -1.3846***   & -1.0269***  \\
              &              & (4.5709)       & (2.9338)     &              & (-9.0643)    & (-5.9488)   \\[1.0em]
Turnover      &              & -0.3199***     & -0.2224***   &              & 1.0092***    & 0.6125**    \\
              &              & (-5.5626)      & (-3.9693)    &              & (4.8642)     & (2.1825)    \\[1.0em]
Age           &              & 0.0001***      & 0.0001       &              & -0.0002***   & -0.0002     \\
              &              & (3.1493)       & (0.0698)     &              & (-4.9529)    & (-0.3302)   \\[1.0em]
Acceleration  &              & 0.1594***      & 0.1613***    &              & -0.5686***   & -0.5767***  \\
              &              & (10.557)       & (10.781)     &              & (-12.267)    & (-12.848)   \\[1.0em]
Intercept     & 0.5183***    & 0.1669***      & 0.2400***    & -0.6695***   & 0.9322***    & 0.5826***   \\
              & (15.389)     & (3.9723)       & (4.8339)     & (-7.1665)    & (8.5946)     & (4.0961)    \\[0.1em]
&  &  &  &  &  &  \\
Observations & 1,017 & 1,017 & 1,017 & 1,017 & 1,017 & 1,017 \\[0.1em]
R-squared             & 0.0700 & 0.2110 & 0.2391 & 0.1263 & 0.3641 & 0.4101 \\[0.25em]
\bottomrule
\end{tabular}}
		\captionsetup{justification=justified, singlelinecheck=on, font=footnotesize}
		\caption{\textbf{Agent-level Crash Predictability.} 
            	The table reports results from cross-sectional linear regressions based on the identified run-up events.
				In specifications (1) to (3), the dependent variable is a dummy indicating an ex-post crash, that is, a cumulative return from $t=1$ to $t=24$ below -40\%.
				In specifications (4) to (6), the dependent variable is the cumulative return from $t=1$ to $t=24$.
				The dependent variables are regressed onto the same set of market-level variables as in Table \ref{tab:fama_reg} and a set of agent-level variables, both measured as of the price run-up identification, at $t=0$. 
				The agent-level variables include:
				(i) \emph{Sophisticated}, the ratio between the number of sophisticated investors actively trading the collection in the ex-ant period \prewin, defined in Section \ref{sub:sophisticated}, and the number of total investors actively trading the collection in the same period;
				(ii) \emph{Unique Owners}, the change in the fraction of unique owners divided by the supply of NFT in the collection recorded during the ex-ante window \prewin;
				(iii) \emph{Wash Trading}, the natural logarithm of one plus the total wash-trading volume in ETH identified by the method described in \ref{sub:wash} before $t=0$.
				\tstats
        	}\label{tab:fama_reg_agent}
		\vspace{0.25em}
	\end{table}

        	Table \ref{tab:fama_reg_agent} presents the outcomes of the aforementioned exercise, indicating that agent-level variables considerably enhance the predictive power of the model in terms of forecasting both crashes and ex-post returns.
			Specifically, three primary observations emerge from the results: 
			First, the presence of sophisticated investors during the ex-ante period is a negative indicator of ex-post crashes. Similarly, the fraction of sophisticated agents positively predicts ex-post returns, exhibiting high levels of statistical significance.
			Second, the change in unique holders exhibits a negative association with the ex-post crash probability, albeit with a lower level of statistical significance. The relation with future returns has a positive point estimate, though it is not statistically distinct from zero.
			Third, a higher quantity of wash trading in the period prior to the run-up identification date is correlated with an increased crash probability and lower ex-post returns. These associations are highly statistically significant in all specifications.
			Taken together, the three additional agent-level predictors elevate the predictive \textit{R}-squared coefficient by nearly $3\%$ for the prediction of crashes and almost $5\%$ for the forecast of ex-post returns, thus increasing the predictive power of the model by more than $10\%$. As we will document in Section \ref{sub:economic_value}, this constitutes an economically significant prediction improvement.
        	
        	These results indicate a correlation between the probability of crashes following price run-ups and different genuine (or nonmanipulative) sophisticated investors, but they do not imply a causal relation. 
        	One may conjecture that sophisticated agents are able to estimate ex-ante the probability of ex-post crashes and, thus, self-select into non-crash events. Although additional analyses are needed to establish causality, the results reported here are consistent with the asset bubble narrative that the predominance of pure speculation is a major factor in the creation of bubbles destined to burst.
        	Further, these findings suggest that wash trading may play a role in the formation of bubbles in the NFT market.
        	A high amount of wash-trading volume may increase the collection's price artificially, thus fostering the formation of a price run-up and increasing the probability of a subsequent crash.

			We finally rerun the full agent-level analysis by excluding the $33,293$ identified wash-trading transactions.
			Our purpose for this exclusion is to assess the impact on the computation of profits for each agent and, consequently, the identification of sophisticated investors. After conducting this analysis, we find that our results remain virtually unchanged and exhibit a slight increase in statistical significance. Based on these findings, we conclude that wash trading does not significantly influence our results, but rather contributes only a minimal amount of noise to our data set.

	    \subsection{Economic Value of Agent-level Variables}\label{sub:economic_value}
			\figurecap{Out-of-sample performance}{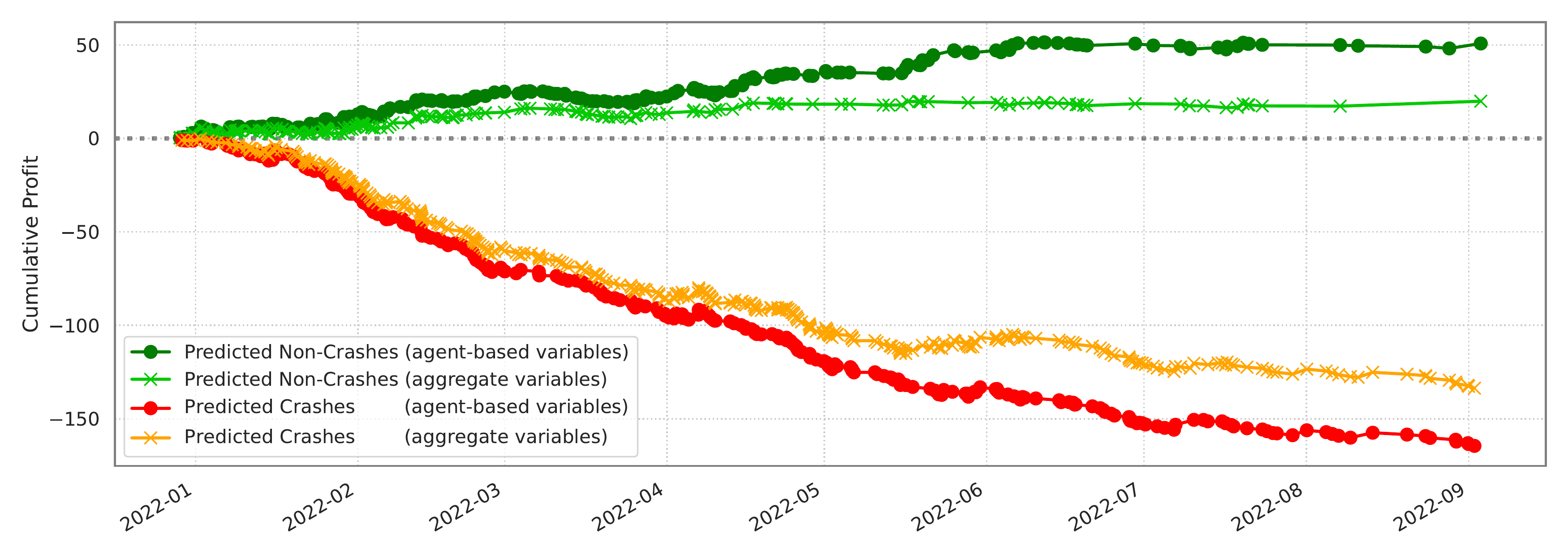}{0.95}{
	            The figure displays the out-of-sample cumulative returns of two portfolios, constructed using a prediction model estimated with agent-level and market-level variables (red and green lines),
	    		and two portfolios constructed using a prediction model estimated with market-level variables only (orange and light green lines).
	    		The prediction models are estimated using data from the $478$ run-up events realized in 2021, while the displayed portfolio returns are based on the $539$ run-up events realized in 2022.
	    		The returns of the ``Predicted Noncrashes" portfolios result from buying into a collection at time $t=1$ when the model prediction is \textit{below} the median of the estimation sample predictions, and selling at $t=24$. 
	    		Similarly, the returns of the ``Predicted Crashes" portfolios result from buying into a collection at time $t=1$ when the model prediction is \textit{above} the median of the estimation sample predictions, and selling at $t=24$. 
	    		Entry and exit prices are based on average sale prices for the collection in the relevant hour.
	        }
	    	To economically assess the level of predictability and, in particular, the value of agent-level variables,
	    	we propose an out-of-sample simple trading strategy based on our regression results.
	    	More precisely, we fit regression models similar to those displayed in columns (2) and (3) of Table \ref{tab:fama_reg_agent}, 
	    	but using only data from run-up events realized in 2021. 
	    	These account for $478$ events, about half of the sample.
	    	We then use the fitted models to produce predictions on the likelihood of an ex-post crash 
	    	for the remaining $539$ events in the sample, those recorded in 2022,
	    	and we construct two calendar-time portfolios based on these predictions.
	    	For the first portfolio, when the predicted crash probability is below the median value of the 2021 sample, 
	    	1 ETH is invested at time $t=1$, and the position is liquidated at the end of the ex-post window ($t=24$). 
	    	Similarly, for the second portfolio, we invest when the predicted crash probability is \emph{above} the median.
	    	The profits and losses are computed based on average traded prices. It should be emphasized that these portfolio strategies are not fully implementable and tradable given the impossibility of short selling.
	    	Figure \ref{fig:strategy_returns.pdf} presents the results from the exercise, showing the performance of the two portfolios. The orange and lightgreen lines represent the time evolution of the economic value of predictions based \emph{only} on market-level variables. The red and green lines are based on a prediction model estimated with  market-level variables \emph{and} agent-level predictors.
            The plot shows that the ex-ante predictability of crashes following price run-ups can generate economically significant profits, 
	        primarily when agent-level variables are used for the estimation of the prediction model.
	        Using only aggregate variables,
	        the portfolio of predicted non-crashes generates a profit of 19 ETH, 
	        while the portfolio of predicted crashes loses 133 ETH.
	        The wedge between the two portfolios significantly increases when we also exploit agent-level variables,
	        generating a profit of 50 ETH and a loss of 164 ETH, respectively.

	\section{Conclusion}\label{sec:conclusion}
		This paper presents a comprehensive study of nonfungible tokens (NFTs), which are unique digital assets recorded on the blockchain, providing proof of ownership and authentication. After constructing a representative sample of the total NFT collection market, we systematically analyze all price run-up events and those that lead to ex-post crashes.
        
        The primary contributions of this research are twofold: first, establishing the predictability of asset bubbles and, second, analyzing the behavior of individual retail investors during asset bubbles. The availability of blockchain technology and the transparency of the NFT market allow us to study these phenomena in a unique way.
		Our analysis reveals that agent-level variables, such as investor sophistication and heterogeneity, in conjunction with asset-specific variables, such as volatility, price acceleration, and turnover, significantly predict bubble formation and price crashes. Sophisticated investors consistently outperform their peers, with indications that they possess superior information and skills, which aligns with existing narratives surrounding asset pricing bubbles.

  	\hypersetup{
        linkcolor={electric_blue},
        citecolor={electric_blue},
        urlcolor={dark_grey}
    }

	\nocite{*}
	\vspace{5em}
	\renewcommand\baselinestretch{1.0}
	\small
	\bibliography{Biblio}

\end{document}